\documentclass[preprint]{ptephy}

\preprintnumber{1410-030} 

\usepackage{ccaption}
\usepackage{bm}

\begin{document}

\title{Effects of  Triple-$\alpha$ and $^{12}\rm C(\alpha,\gamma)^{16}O$ Reaction Rates
on the Supernova Nucleosynthesis in a Massive Star of 25 $M_{\odot}$}

\author{\name{Yukihiro \textsc{Kikuchi}}{1,\ast}, \name{Masa-aki \textsc{Hashimoto}}{1,\ast}, \name{Masaomi \textsc{Ono}}{1}, and \name{Ryohei \textsc{Fukuda}}{1}
}

\address{\affil{1}{Department of Physics, Kyushu University, Fukuoka 812-8581, Japan}
\email{kikuchi@phys.kyushu-u.ac.jp, hashimoto@phys.kyushu-u.ac.jp}}
\begin{abstract}%
We investigate effects of triple-$\alpha$ and $^{12}\rm C(\alpha,\gamma) ^{16}O$ reaction rates on the production of supernova yields for a massive star of 25 $M_{\odot}$. 
We combine the reaction rates to examine the rate dependence, where the rates are considered to cover the possible variation of the rates based on experiments on the earth and theories. We adopt four combinations of  the reaction rates from two triple-$\alpha$ reaction rates and two $^{12}\rm C(\alpha,\gamma)^{16}O$ ones. First, we examine the evolution of massive stars of  20 and 25  $M_{\odot}$ whose helium cores correspond to helium stars of 6 and 8 $M_{\odot}$, 
respectively. While the 25 $M_{\odot}$ stars evolve to the presupernova stages for all combinations of 
the reaction rates,  evolutionary paths of the 20 $M_{\odot}$ stars proceed significantly different way for some combinations, which are unacceptable for progenitors of supernovae.
Second,  we perform calculations of supernova explosions within the limitation of spherical symmetry and
compare the calculated abundance ratios with the solar system abundances. We can deduce some constraints to the reaction rates.
As the results, a conventional rate is adequate for a triple-$\alpha$ reaction rate and a rather higher value of the 
reaction rate within the upper limit for the experimental uncertainties 
is favorable for a $^{12}\rm C(\alpha,\gamma)^{16}O$ rate. 
\end{abstract}

\subjectindex{D40, E25, E26}

\maketitle

\section{Introduction}

\begin{figure}[h!]
\begin{center}
\includegraphics[scale=1.0]{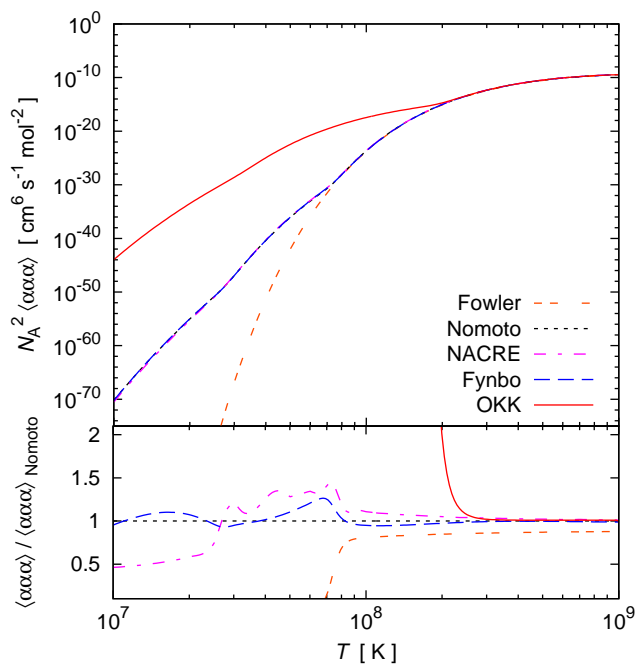} 
\vspace{1.5cm}
\caption{Triple-$\alpha$ reaction rates as a function of temperature. `Fowler', `Nomoto', `NACRE', `Fynbo' and `OKK' are 
the 3$\alpha$ reaction rates 
taken from Fowler et al. (1975) \cite{Fowler1975}, Nomoto et al. (1982a) \cite{rf:ntm85}, NACRE (Angulo et al. 1999) \cite{rf:angulo}, Fynbo et al. (2006) \cite{rf:fynbo}, and Ogata et al. (2009) \cite{rf:okk}, respectively. Top panel shows thermonuclear reaction rates and lower panel shows the ratios with respect to `Nomoto'. $\langle \alpha \alpha \alpha \rangle$ is the reaction rate per three $\alpha$ particles.}
\label{fig:rate1}
\end{center} 
\end{figure}

The triple-$\alpha$ (3$\alpha$) and $^{12}\rm C(\alpha,\gamma)^{16}O$ reactions play a crucial role in helium 
(He) and carbon (C) burning stages on the evolution of low-, intermediate-, high-mass stars \cite{rf:sn80,rf:nh88,rf:hashi95}, and accreting compact stars \cite{rf:nom82,rf:nom82b,rf:ntm85}.
 The 3$\alpha$ reaction rate  calculated by Ogata et al. \cite{rf:okk} (hereafter we call the rate as OKK rate) is very large compared with previous rates \cite{rf:ntm85,rf:angulo} for temperature below $2 \times10^8$~K.
The rate by Fynbo et al.~\cite{rf:fynbo} (Fynbo rate) which revised the 3$\alpha$ rate of Ref.~\cite{rf:angulo} is based on new experiments at high temperature of $T > 10^{9}$~K. 
The ratios of the OKK rate to the Fynbo rate are 1.0, 1.9, $1.5 \times 10^{6}$, and $3.6 \times 10^{26}$ for $T_{8}$ = 2.5, 2, 1, and 0.1, respectively ($T_{8}$ is temperature in units of $10^8$~K). 
The OKK rate is shown in Fig.~\ref{fig:rate1}, where the representative rates obtained so far are compared. 
It is noted that a conventional formula 
$\dot{y}_{12}=-1/6\rho^2 N_A^2\langle \alpha \alpha \alpha \rangle y_{4}^{3}$  with the abundance $y_i = n_i/(\rho N_A)$ 
is used, where $n_i$ is the number density of a nucleus $i$  ($i=12$ for $^{12}\rm C$ and  $i=4$ for $^{4}\rm He$), $\rho$ is the mass density, $N_{\rm A}$ is the Avogadro constant~\cite{rf:ntm85}.
We must investigate how the new rates affect the astrophysical phenomena, because terrestrial experiments for the 3$\alpha$ reaction are very difficult. 

\begin{figure}[b!]
\begin{center}
\includegraphics[scale=1.0]{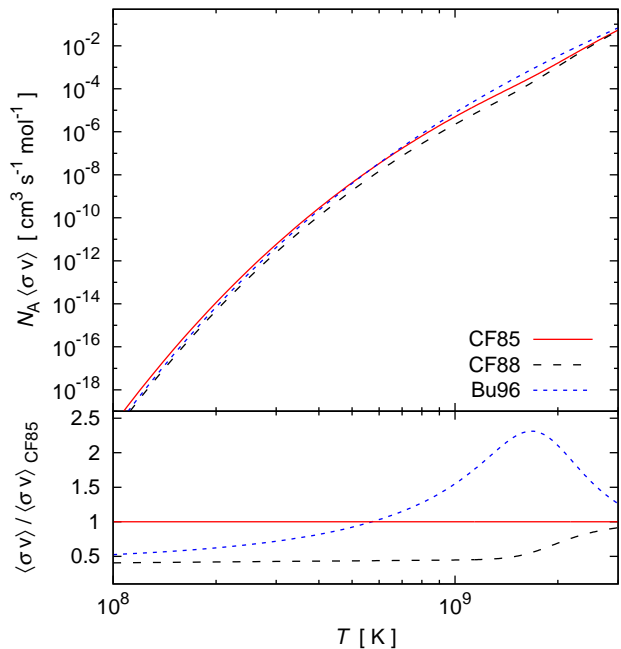} 
\end{center} 
\vspace{1.5cm}
\caption{Thermonuclear reaction rates for $^{12}$C($\alpha,\gamma$)$^{16}$O as a function of temperature. The abbreviations `CF85', `CF88', and `Bu96' are taken from Caughlan et al. (1985) \cite{cfhz85}, Caughlan and Fowler (1988) \cite{cf88}, and Buchmann et al. (1996a) \cite{bu96}, respectively. Upper panel shows thermonuclear reaction rates per particle and lower panel shows the ratios with respect to `CF85'.}
\label{fig:rate2}
\end{figure}

The uncertainty of the rate of the $\rm^{12}C(\alpha,\gamma)^{16}O$ is known to be within a factor of two suspected from nuclear experiments \cite{cfhz85,cf88}.
The direct measurements of this reaction are very difficult for the low energy region 
corresponding to the stellar energy of 300 keV at which we need the cross section. 
The astrophysical $S$-factor at this energy, $S(300)$, is estimated to be 100 and 230~keV$\cdot$b from Refs.~\cite{cf88} and \cite{cfhz85}, respectively. 
In particular, the latter rate, referred to as CF85, has been adopted to explore supernova nucleosynthesis and the final results seem to give a good agreement with the observation of SN 1987A~\cite{rf:hashi95}.  
Another value has been presented with use of a different method to determine the cross section, which utilizes the reacion of
$\rm ^{16}N(\beta^-)^{16}O^* \to ^{12}C + \alpha$  and gives $S(300) = 146$~keV$\cdot$b (hereafter referred to as Bu96) \cite{bu96}.
Although the uncertainty is not so large compared to the triple-$\alpha$ rate 
as shown in Fig.~\ref{fig:rate2}, where the representative rates  obtained so far are compared, the effects on the nucleosynthesis 
is significant \cite{tur2007,tur2009,austin2014,parikh2013}. 

 It has already been shown that the OKK rate crucially affects the evolutionary tracks of low-mass stars, where the evolution from  zero-age main sequence to core He flash/burning for low-, intermediate-, and high-mass stars have been investigated \cite{rf:dotter,rf:morel}. 
The HR diagram obtained using the new 3$\alpha$ reaction rate disagrees considerably with the observations of low-mass stars; 
the OKK rate results in the shortening or disappearance of the red giant phase, because helium ignites at much lower temperature and density compared to the case of the NACRE rate~\cite{rf:angulo}. 
Furthermore, stellar models  in the mass range of $0.8< M/M_{\odot}< 25$ were computed and it was confirmed that
the OKK rate has significant effects on the evolution of low- and intermediate-mass stars, while its influences on the 
evolution of massive stars ($M > 10 M_{\odot}$) are minimal~\cite{suda}; the OKK  rate is  incompatible with 
observations but for massive stars. If the OKK rate is correct, we must invoke some new physical processes such as 
rotational mixing~\cite{mm97,maeder97}, turbulence~\cite{meakin+arnett07}, 
dynamical instabilities~\cite{smith+arnett14}
or other unknown physical effects. 
On the other hand, the abundances of  helium and heavier elements in globular clusters are open to dispute \cite{rf:piot07}, which may change the scenario of the stellar evolution of low mass stars. 
Apart from appearances of observations, we can see the effects of the OKK rate on stellar evolution from the ignition properties. A helium core  flash is triggered if the nuclear energy generation rates ($\varepsilon_{\rm n}$) become significantly larger than the neutrino energy loss rates ($\varepsilon_{\nu}$). 
We can understand clearly that helium ignition under the degenerate condition ($\varepsilon_{\rm n} = \varepsilon_{\nu}$) 
occurs at considerably low temperature and density points compared with the previous case \cite{saru2010}. 
The effects of the OKK rate on the evolution of accreting compact stars have been studied: the ignition property for accreting white dwarfs~\cite{saru2010} and X-ray bursts on accreting neutron stars \cite{matsuo2011}. 
It was also found that  the $s$-process with use of the OKK rate during core He-burning is very inefficient compared to the case 
with the previous 3$\alpha$ rates. However, the difference of the overproduction is found to be almost compensated by the subsequent 
C-burning and 
the overproduction level is not different as a whole for the two distinctly different 3$\alpha$ rates. Therefore, the weak $s$-process in massive stars 
does not testify the validity of the new rate. 
Tur et al.~\cite{tur2009} investigated the dependence of the $s$-process and 
post-explosive nuclosynthesis on $\pm2 \sigma$ experimental uncertainties of 3$\alpha$ and $^{12}\rm C(\alpha,\gamma)^{16}\rm O$ reaction rates. However, the impact of 
such a large theoretical uncertainties of 3$\alpha$ rates invoked by 
OKK rate~\cite{rf:okk} with those of $^{12}\rm C(\alpha,\gamma)^{16}\rm O$ 
has not been explored on the supernova yields of a massive star.

In the present paper, we investigate the effects of both 3$\alpha$ and $^{12}\rm C(\alpha,\gamma)^{16}\rm O$ rates on the production of the possible isotopes during the evolution of  a massive star of 25 $M_{\odot}$ and 
its supernova explosion. 
In \S2, the evolution of massive stars of 20 $M_{\odot}$ and 25 $M_{\odot}$ stars are presented, where the effects of the rates on the evolution and the nucleosynthesis are discussed. The method of calculations of the supernova explosion of the 25 $M_{\odot}$ stars 
is presented in \S3. The results of nucleosynthesis and some discussions are also given by comparing with the solar system abundances. 
In \S4, the most suitable combination of the reaction rates are deduced and remaining problems are given.

\section{Nucleosynthesis at the presupernova stages}

3$\alpha$ and $^{12}\rm C(\alpha,\gamma)^{16}O$  rates are the key nuclear reaciton rates concerning He-burning in the massive star evolution.  
As a consequence, explosive nucleosynthesis and
resulting supernova yields of a massive star would be influenced seriously by the two rates.
We select four combinations from the available nuclear data for  the two rates, that is, Fynbo-CF85, Fynbo-Bu96, OKK-CF85, and OKK-Bu96, which must cover the possible uncertainties inherent to the experiments and/or theories. The stellar evolutionary code is almost the same as Ref.~\cite{rf:nh88,rf:hashi95} but for the revised reaction rates \cite{cf88}.
To study  detailed abundance distributions  including $s$-nuclei, we perform a post-process nucleosynthesis calculation with a large nuclear reaction network using the same methods as described in Ono et al.~\cite{ono2012} and Kikuchi et al.~\cite{kiku2012}. 

Let us explain our nuclear reaction network for completeness.
Our network  contains 1714 nuclei from neutron and proton to uranium isotopes up to 
$^{241}$U linked through particle reactions and weak interactions~\cite{nishi06,ono2012}.
The reaction rates are taken from JINA REACLIB compilation~\cite{cyburt}, where  updated nuclear data has been included for the charged particle reactions and ${\rm (n,\gamma)}$ cross sections after those of Bao et al.~\cite{bao00}. 
The finite temperature and density dependences of beta-decay and electron capture rates for nuclei above 
$^{59}$Fe are included based on the Ref.~\cite{ty87}.

After helium core formation, gravitational contraction leads to the ignition by the 3$\alpha$ reaction. 
Near the end of core He-burning, the $^{12}\rm C(\alpha,\gamma)^{16}O$ reaction  begins to operate significantly. 
As a consequence, the production of $^{12}$C and $^{16}$O proceeds appreciably and these elements should become main products after core He-burning in all massive stars. 
>From core C-burning to the end of core oxygen burning, carbon continues to decrease little by little due to shell burnings.
This fact is almost universal from zero to solar metallicity stars, because massive stars form helium cores after hydrogen burning except for extremely massive stars which could induce pair instability supernovae \cite{un2002}.

\begin{figure}
\begin{center}
\hspace*{-1cm}
{\includegraphics[scale=1.3]{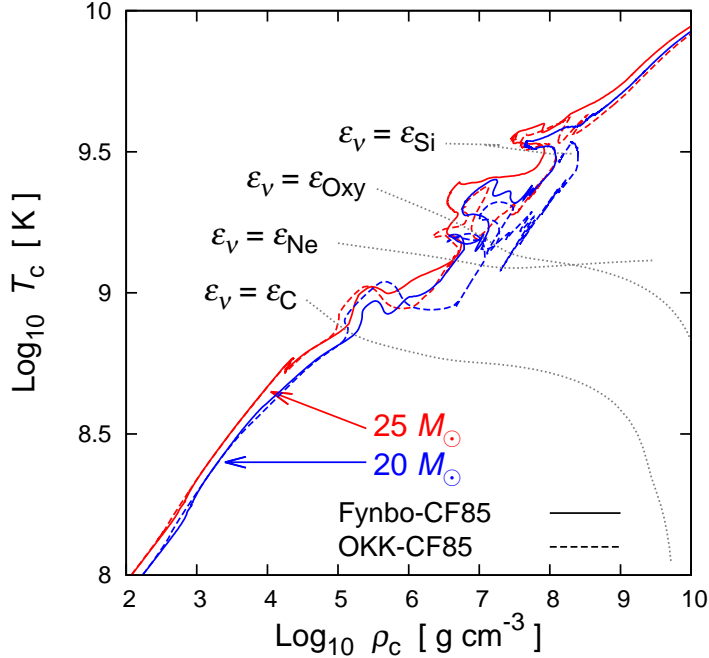}} 
\end{center}
\caption{Evolutionary paths  on the plane for the density and the temperature at the center of the 20 and 25 $M_\odot$ stars. The solid and dashed lines are the results from Fynbo and OKK rates with use of CF85, respectively. 
The ignition curves (dotted lines) show the beginning of C-, Ne-, O-, and Si-burning, where the neutrino loss rates are equal to the nuclear energy generation rates.}
\label{fig:rhot1}
\end{figure}

\begin{figure}
\begin{center}
\hspace*{-1cm}
\includegraphics[scale=1.3]{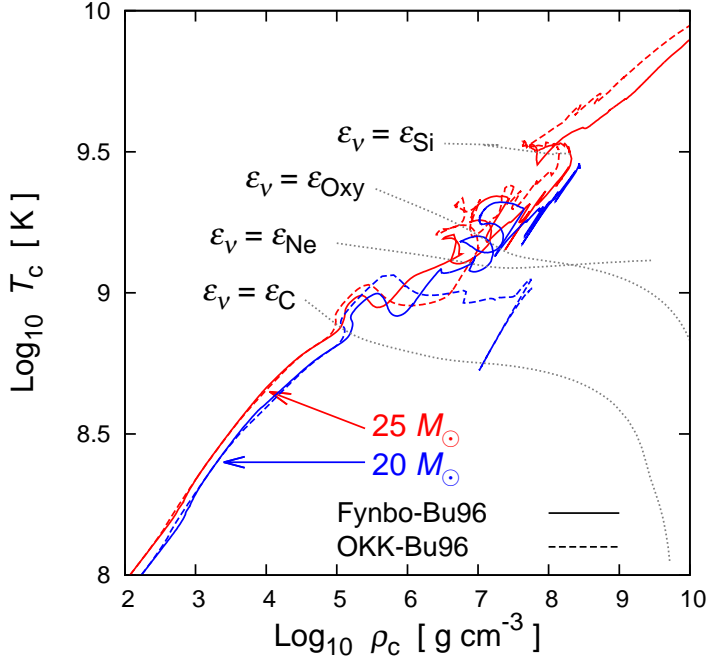} 
\end{center}
\caption{Same as Fig.~\ref{fig:rhot1} except for the reaction rate Bu96 of $^{12}\rm C(\alpha,\gamma)^{16}O$. }
\label{fig:rhot2}
\end{figure}

Figure~\ref{fig:rhot1} shows the evolutionary tracks for the density and temperature at the center in cases of  the 20 and 25 $M_\odot$ stars, where  Fynbo-CF85 and OKK-CF85 rates are adopted. The evolution with respect to the 25 $M_\odot$ stars lead to presupernova
stages for both rates. The situation becomes rather complex if we examine the evolution of the 20 $M_\odot$ stars.  For the case of Fynbo-CF85, the presupernova
stage is attained as before~\cite{rf:nh88,rf:hashi95}. For the case of OKK-CF85, Si-ignition occurs barely and the evolution would lead to the formation of the Fe-core. 
In the present study, we stopped the calculation of Si-burning; since the shell burnings of O, Ne, and C
are often very active, the computation becomes difficult to continue. 
The details of the tracks depend on both the strength of the shell burnings and the extension of the convective mixing. Finally, the presupernova stages are attained as seen in Fig.~\ref{fig:rhot2} for the 25 $M_{\odot}$ stars. 
On the other hand, we find that  the presupernova stage cannot be obtained for 20 $M_\odot$ star with the Fynbo-Bu96 model but instead the star begins to cool as seen in Fig.~\ref{fig:rhot2}; the 
central region cannot attach the ignition curve of Si on the density-temperature plane. 
In general, whether nuclear ignition occurs or not can be judged from the temperature to which the burning region heats up.  
If the temperature does not reach the ignition temperature, the region begins to cool. The central temperature depends on significantly
the strength of shell burnings. Especially, the production of carbon at the end of
core helium burning is closely related to the subsequent evolution of the stars, because carbon shell determines the boundary between
the inner core of carbon-oxygen core and helium envelope. Furthermore, active carbon burning hinders the increase of the central 
temperature and leads to the delay of gravitational contraction. 
%
Therefore, the formation of the Fe-core is doubtful if we adopt the combination 
of the reaction rates of OKK-CF85 or Fynbo-Bu96, because the central temperature 
is just around the ignition line for Si. 
We note that once the ignition of silicon begins, 
Fe-core forms at the center and gradually grows towards the Chandrasekhar 
mass~\cite{rf:nh88}. 
Although the model of OKK-CF85 may not be excluded as a progenitor,
the evolutionary scenario will become complex compared to the previous scenario~\cite{rf:nh88,rf:hashi95}.
It should be noted that the evolution becomes very complex for massive stars whose masses are less than 20 $M_\odot$. This can
be inferred from Figs.~\ref{fig:rhot1} and~\ref{fig:rhot2};  
The reaction rates OKK-Bu96 change drastically the evolutionary history of the 20 $M_{\odot}$ stars 
compared to the case with Fynbo-CF85, 
and the evolutionary path resembles to a star of around 10 $M_\odot$, where the nuclear ignition for neon burning does not occur at the center~\cite{nh86}. 
As a consequence, the evolutionary scenario of less massive stars ($M\leq 20~M_{\odot}$) would change significantly. 
\begin{figure}[!b]
\hspace*{-0.5cm}
\begin{tabular}{cc}
\begin{minipage}{80mm}
\mbox{\raisebox{1mm}{\includegraphics[width=8cm]{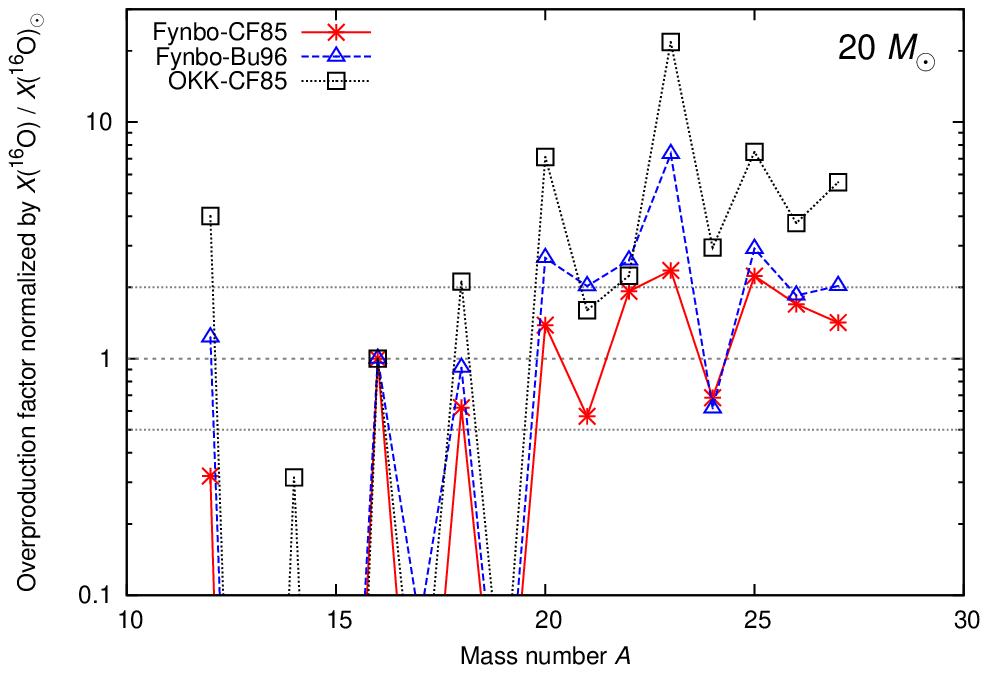}}} 
\caption{{The overproduction factors of the nuclei lighter than $A = 27$ in the 20 $M_{\odot}$ stars at the start of Si-burning, which are normalized by the abundance of $^{16}$O. The asterisks, triangles, and squares are the results from the models of Fynbo-CF85, Fynbo-Bu96, and OKK-CF85, respectively.
The two dotted lines show the values whose ratios to the normalized overproduction 
factors of $^{16}$O are two or one-half.
}}
\label{fig:6Mlight}
\end{minipage}
\mbox{\raisebox{1mm}{
\begin{minipage}{80mm}
\begin{center}
\vspace*{-3cm}
\mbox{\raisebox{1mm}{\includegraphics[width=8cm]{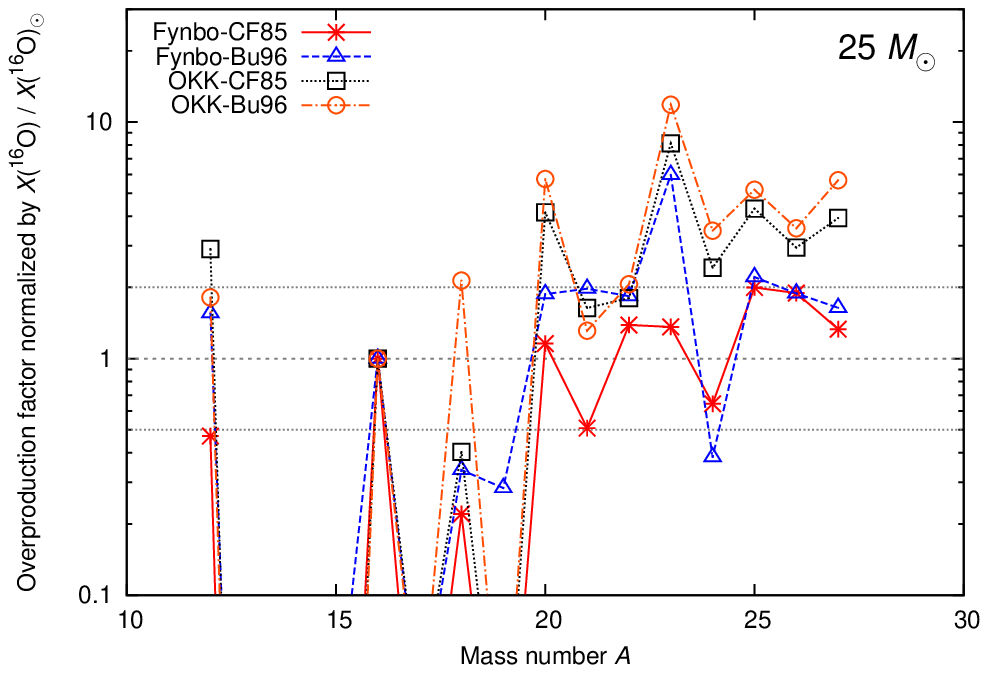}}} 
\caption{{Same as Fig.~4 but for  the 25 $M_{\odot}$ stars at the presupernova stages with use of all 
combinations of reaction rates.}}
\label{fig:8Mlight}
\end{center}
\end{minipage}}}
\end{tabular}
\end{figure}
\begin{figure}
\begin{center}
\begin{tabular}{cc}
\resizebox{15cm}{!}{\includegraphics{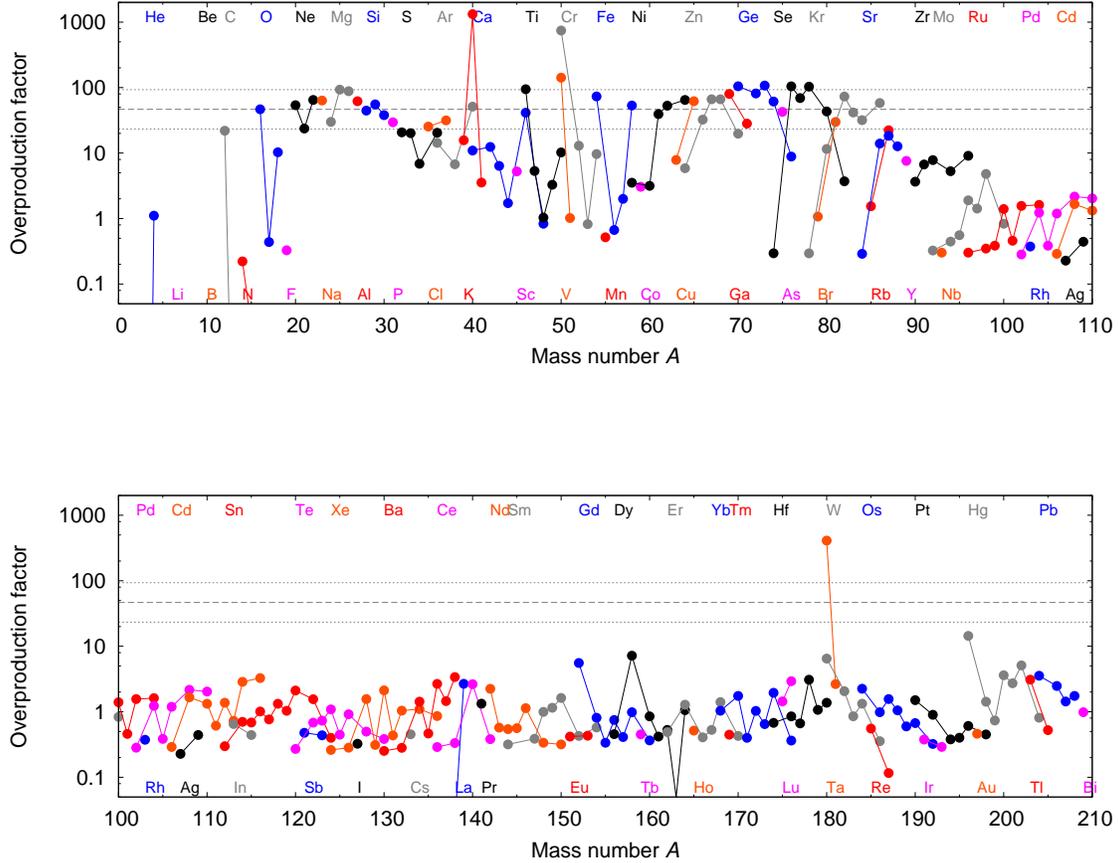}} 
\end{tabular}
\vspace{-1.5cm}
\caption{Overproduction factors of all stable nuclei of $A < 210$ for the Fynbo-CF85 model at the presupernova stage 
for 25 $M_{\odot}$ star. 
The dotted lines denote the values whose ratios to the overproduction factor of $^{16}$O are two or one-half. 
The dashed line shows the value of $^{16}$O.
}
\label{fig:prefycf}
\end{center}
\end{figure}
%

Figures~\ref{fig:6Mlight} and \ref{fig:8Mlight} illustrate the overproduction factors $X(i)/X(i)_{\odot}$, 
where $X(i)$ denotes the mass fraction of element $i$, concerning main products  ($12 \leq A \leq 27$) 
normalized by the oxygen one, $X(^{16}{\rm O})/X(^{16}{\rm O})_{\odot}$. 
The two dotted lines show the values whose ratios to the normalized overproduction 
factor of $^{16}$O are two or one-half. 
Figure~\ref{fig:6Mlight} shows the case of the 20 $M_{\odot}$ stars at the beginning of Si-burning for three combinations of reaction rates; it is noted that the abundances referred in the panel does not change appreciably after Si-burning. 
Figure~\ref{fig:8Mlight} shows the case of the 25 $M_{\odot}$ stars at the presupernova stages for the four combinations. 
As a whole, the Fynbo-CF85 model gives a reasonable range of abundance ratio compared to the solar abundances but for $^{12}$C which is supplied through the ejection from AGB stars~\cite{latter91,gh97}. 
%
%
For other models, 
however, the overproduction factors of some
elements are beyond the values whose ratios to
those of oxygen are two (upper dotted lines). 

In Fig.~\ref{fig:6Mlight} and Fig.\ref{fig:8Mlight},  the models except for Fynbo-CF85 produce a significantly large amount of $^{20}$Ne, $^{23}$Na, and $^{24}$Mg because of strong C-burning. 
Since these elements remain in the burning layers of O and Ne, significant amounts will survive even after the shock wave propagation during the explosion. The amounts can be obtained by simulating the explosion. 
The fate of  a star towards Fe-core formation is still uncertain due to  convective mixing even under the assumption 
of spherical symmetry~\cite{rf:nh88,rf:hashi95}.
However, we have succeeded in the stellar evolutionary calculations until the beginning of Fe-core collapse only for 25 $M_{\odot}$ star for the four 
reaction sets as seen in Figs.~\ref{fig:rhot1} and~\ref{fig:rhot2}. 
It is desirous to adopt the complete set of presupernova models to examine the effects of four sets of reaction rates on nucleosynthesis. 
Therefore, we show the nucleosynthesis after the supernova explosions in the next section by concentrating on the 25~$M_{\odot}$ stars. 

In Kikuchi et al.~\cite{kiku2012}, 
we have already investigated the effects of the OKK rate on the production of the $s$-process nucleosynthesis for a 25~$M_{\odot}$ star. 
However, the calculation was stopped at the end of central C-burning.  
Therefore, we can discuss the $s$-process nucleosynthesis after 
central C-burning in the present paper. Overall, at the end of C-burning, 
the overproduction factors of $s$-nuclei are roughly consistent 
with those of Kikuchi et al.~\cite{kiku2012}. After C-burning, neutron irradiations in 
C-burning shells enhance the overproduction factors of some $s$-nuclei 
by a factor of 3--4 (for $^{80}$Kr, even greater than 10) 
compared to those at the end of C-burning. The significant enhancement of the $s$-process elements during the later evolutionary stages has already been claimed by Tur et al.~\cite{tur2009}. We have 
confirmed that the level of the enhancement of overproduction factors due to 
the later burning stages is roughly consistent with that in the Ref.~\cite{tur2009}.
Since $s$-elements become seeds of $p$-process nucleosynthesis in massive stars and
the $p$-process has been believed to occur during a 
supernova explosion at the bottom of the oxygen-rich layer, how much amount of $s$-elements can be survived is crucial to
discuss the nucleosynthesis of $p$-elements in supernovae. 

Let us examine the nucleosynthesis
for the nuclei above $A>40$ other than $s$-nuclei concerning the 25 $M_{\odot}$ stars. 
In Fig.~\ref{fig:prefycf}, the Fynbo-CF85 model gives significant overproductions compared to the initial solar abundances for the
neutron rich nuclei of 
$^{40}$K (1325), $^{50}$V (142), $^{50}$Cr (744), and $^{180}$Ta (410), 
where the numerals inside the brackets are the overproduction factors with respect to their  initial values. 
In the present case, we summed the mass of each nucleus above approximately  1.5~$M_{\odot}$ which corresponds to a `mass cut' described in \S3. 
On the other hand, other models do not produce much of those nuclei except 
for $^{40}$K, where overproduction factors are 1092, 1347, and 1281 for Fynbo-Bu96, OKK-Bu96, and OKK-CF85, respectively. 
For other nuclei, three models result in the overproduction factors less than 100. 
Exceptionally, Fynbo-Bu96 and OKK-Bu96 models give overproduction factors 
of 142 for $^{76}$Se and  141 for $^{50}$Cr, respectively. 
It is noted that the first three nuclei ($^{40}$K, $^{50}$V, $^{50}$Cr) are the products after the oxygen burning. As seen in Fig.~\ref{fig:8Mlight}, oxygen production
overwhelmed carbon production after He-burning for the Fynbo-CF85 model. 
As a consequence, the model  tends to well overproduce these nuclei as can be seen in Fig.~\ref{fig:prefycf}. $^{180}$Ta is mainly produced by ($\gamma$,$\,$n) reaction 
of $^{181}$Ta. The overproduction of $^{180}$Ta for Fynbo-CF85 is attributed to the 
enhanced ($\gamma$,$\,$n) reaction owing to higher temperatures during 
the later evolutionary stages as seen in Fig.~\ref{fig:rhot1}. 
After the supernova explosion, these nuclei around the bottom of oxygen-rich layers are destroyed and/or transformed to other nuclei.
Therefore, except for some nuclei, overproductions are decreased as is discussed in the next section.
In the following section, we focus on the nucleosynthesis of 25~$M_{\odot}$ stars, because we can succeed in getting the presupernova models
for the four models of Fynbo-CF85, Fynbo-Bu96, OKK-CF85, and OKK-Bu96. The more massive stars will result in the straight forward
evolution toward the Fe-core collapse. The less massive stars experience rather complex evolutions as inferred from  Fig.~\ref{fig:rhot2}.

\section{Supernova nucleosynthesis and overproduction factors}

We investigate the production of the elements for a massive stars of 25 $M_{\odot}$. 
To estimate the amount of ejected materials into the interstellar medium from the exploding star, we perform the simulation of the supernova explosion.
The procedure of this calculation has been described in the preceding study \cite{ha1989}.
Therefore, we explain briefly the calculation method.
The equations of hydrodynamics are as follows with use of the Lagrange mass coordinate $m$ (e.g., Ref~\cite{yahili});
\begin{align}
\label{eq:ppm1} \frac{\partial V}{\partial t} - \frac{\partial (4\pi r^2 v)}{\partial m} &= 0, \\
\label{eq:ppm2}\frac{\partial q}{\partial t} + \frac{\partial (4\pi r^3 p)}{\partial m} &= v^2 + \frac{3p}{\rho} -\frac{Gm}{r}, \\
\label{eq:ppm3}\frac{\partial e}{\partial t} + \frac{\partial (4\pi r^2 v p)}{\partial m} &= H.
\end{align}
Eq. \eqref{eq:ppm1} describes the equation of continuity with the specific volume $V=\rho^{-1}$, where $r$ and $v$ are the radius and velocity, respectively. 
Eq. \eqref{eq:ppm2} is the conservation of momentum, here $p$ is the pressure, $q$ is the scalar specific momentum described as $q = \bm{v}\cdot\bm{r}$, and $G$ is the gravitational constant.
Eq. \eqref{eq:ppm3} gives the equation of energy conservation; $e$ is the specific energy expressed as $e=\frac{v^2}{2}+U-\frac{Gm}{r}$, where $U$ is the specific internal energy, and $H$ is the heating term by the nuclear energy generation (energy per unit mass per unit time). 

Concerning the initial models, we adopt presupernova models which are obtained in \S2. 
Input physical values are the temperature, density, pressure and chemical compositions. Our hydrodynamical code includes $\alpha$-network which contains 13 species: $^4$He, $^{12}$C, $^{16}$O, $^{20}$Ne, $^{24}$Mg, $^{28}$Si, $^{32}$S, $^{36}$Ar, $^{40}$Ca, $^{44}$Ti, $^{48}$Cr, $^{52}$Fe and $^{56}$Ni~\cite{snh88}. 
The calculation continues until the time of around 300~s when the shock wave reaches the surface of the helium 
core; at the same time the explosion energy is calculated.
The explosion is initiated by injecting a thermal energy around the surface of 
the Fe-core. To see the effects of different combinations of the reaction rates, we 
fix the explosion energy and ejected $^{56}$Ni mass to be $1.0 \times10^{51}$~erg 
and 0.07 $M_{\odot}$, respectively. We adopt these values from SN 1987A~\cite{snh88} 
as a core-collapse supernova explosion model. The injected energy is adjusted 
to obtain an explosion energy of $1.0 \times10^{51}$~erg. 
We note that the locations in Lagrange mass coordinate 
at which the thermal energies are injected, i.e. the surfaces of the Fe-cores, 
are different for each model, 
because different combination of the reaction rates result in different Fe-core masses. 
After the nucleosynthesis calculation described later, we redifine a 
boundary between the ejecta and the compact object which is 
so-called a `mass-cut' ($M_{\rm cut}$) to obtain 0.07 $M_{\odot}$ $^{56}$Ni in 
the ejecta. It is assumed that the material between $M_{\rm Fe}$ and $ M_{\rm cut}$ 
($M_{\rm Fe} < M_{\rm cut}$) falls into the compact object. 
In Table I, the physical quantities concerning the explosion, that is, 
the mass of the Fe-core $M_{\rm Fe}$, the injected energy $E_{\rm in}$, and the mass cut $M_{\rm cut}$ are shown. 

Using the results of the density and temperature evolution during the shock propagation, we calculate nucleosynthesis
 with a large nuclear reaction network.
The calculations are performed until $10^{17}$ s after the explosion, which leads to stable nuclei (we extrapolate the density and temperature after 300 s to continue the nucleosynthesis calculation assuming an adiabatic expansion). The reaction network is almost the same as that of the evolution calculation with 1714 species but we add proton rich elements around Fe group nuclei for explosive nucleosynthesis, whose network includes 1852 nuclear species.
%
\begin{table}[h]
\begin{center}
\label{table:input parameter}
\caption{Physical quantities for the explosion models. The Fe-core mass $M_{\rm Fe}$ and mass cut $M_{\rm cut}$ are 
the values in units of $M_{\odot}$. The injected energy $E_{\rm in}$ is in units of 10$^{51}$ erg.}\vspace{1eM}
\begin{tabular}{ccccc} \hline
\hspace{0ex} Models       & \hspace{0ex} Fynbo-CF85 & \hspace{0ex} Fynbo-Bu96 & \hspace{0ex} OKK-CF85 & \hspace{0ex} OKK-Bu96  \hspace{0ex} \\ \hline
\hspace{0ex} $M_{\rm Fe}$ & \hspace{0ex} 1.36 & \hspace{0ex} 1.25 & \hspace{0ex} 1.24 & \hspace{0ex} 1.34 \hspace{0ex} \\ 
\hspace{0ex} $E_{\rm in}$ & \hspace{0ex}  1.44 & \hspace{0ex} 1.06 & \hspace{0ex} 1.30 & \hspace{0ex} 1.16 \hspace{0ex} \\
\hspace{0ex} $M_{\rm cut}$ & \hspace{0ex} 1.70 & \hspace{0ex} 1.35 & \hspace{0ex} 1.46 & \hspace{0ex} 1.53 \hspace{0ex} \\ \hline
\end{tabular}
\end{center}
\end{table}

\begin{figure}
\begin{center}
\begin{tabular}{cc}
\resizebox{15cm}{!}{\includegraphics{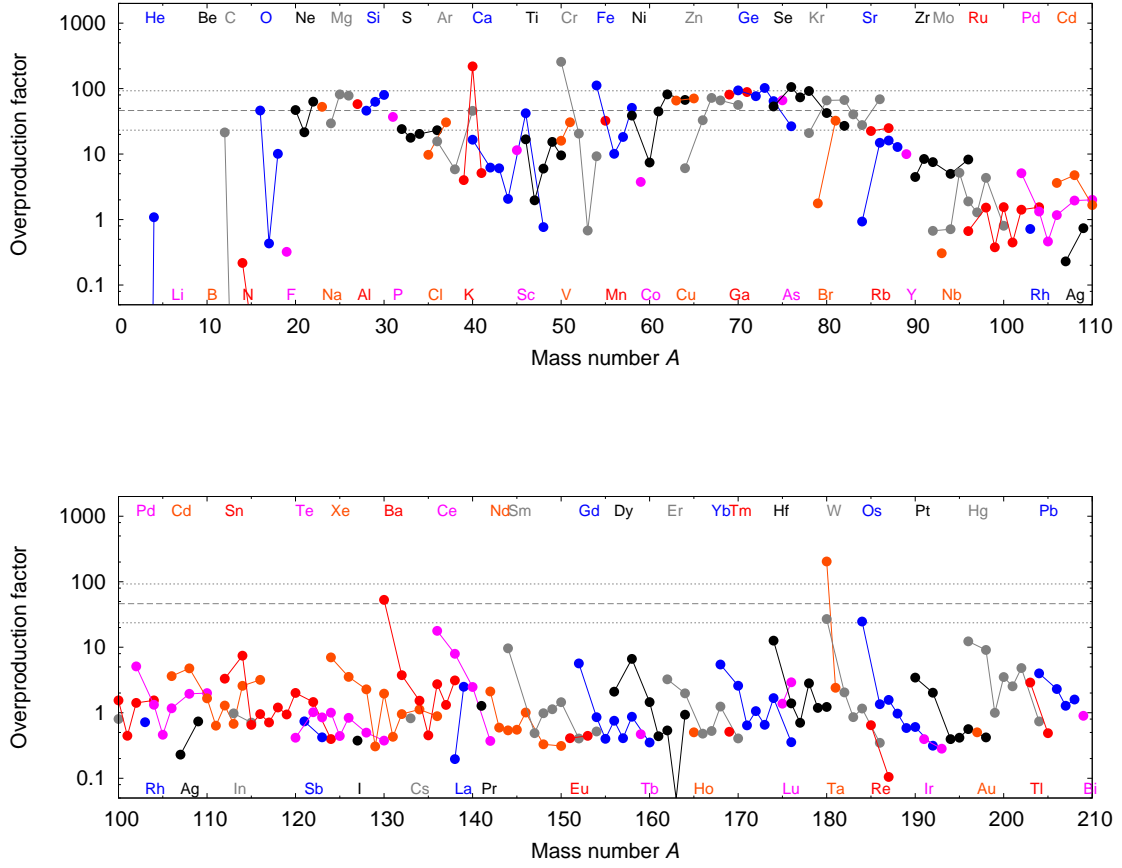}} 
\end{tabular}
\vspace{-1.5cm}
\caption{Overproduction factors of all stable nuclei of $A < 210$ for Fynbo-CF85 model at $10^{17}$ s after the 
explosion. 
The dotted lines denote the values whose ratios to the overproduction factor of $^{16}$O are two or one-half. 
The dashed line shows the value of $^{16}$O.
}
\label{fig:all_fycf}
\end{center}
\end{figure}

\begin{figure}
\begin{center}
\begin{tabular}{cc}
\resizebox{15cm}{!}{\includegraphics{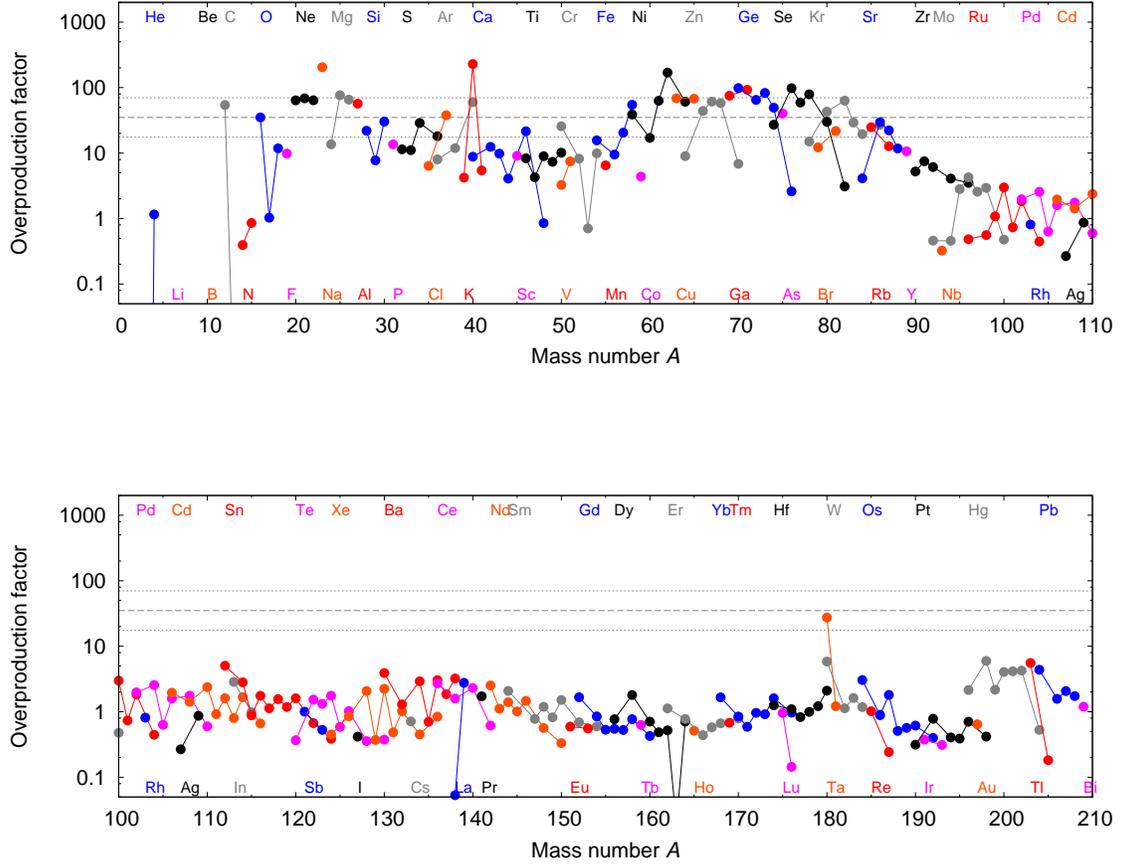}} 
\end{tabular}
\vspace{-1.5cm}
\caption{Same as Fig.~\ref{fig:all_fycf} but for the Fynbo-Bu96 model.}
\label{fig:all_fybu}
\end{center}
\end{figure}

\begin{figure}
\begin{center}
\begin{tabular}{cc}
\resizebox{15cm}{!}{\includegraphics{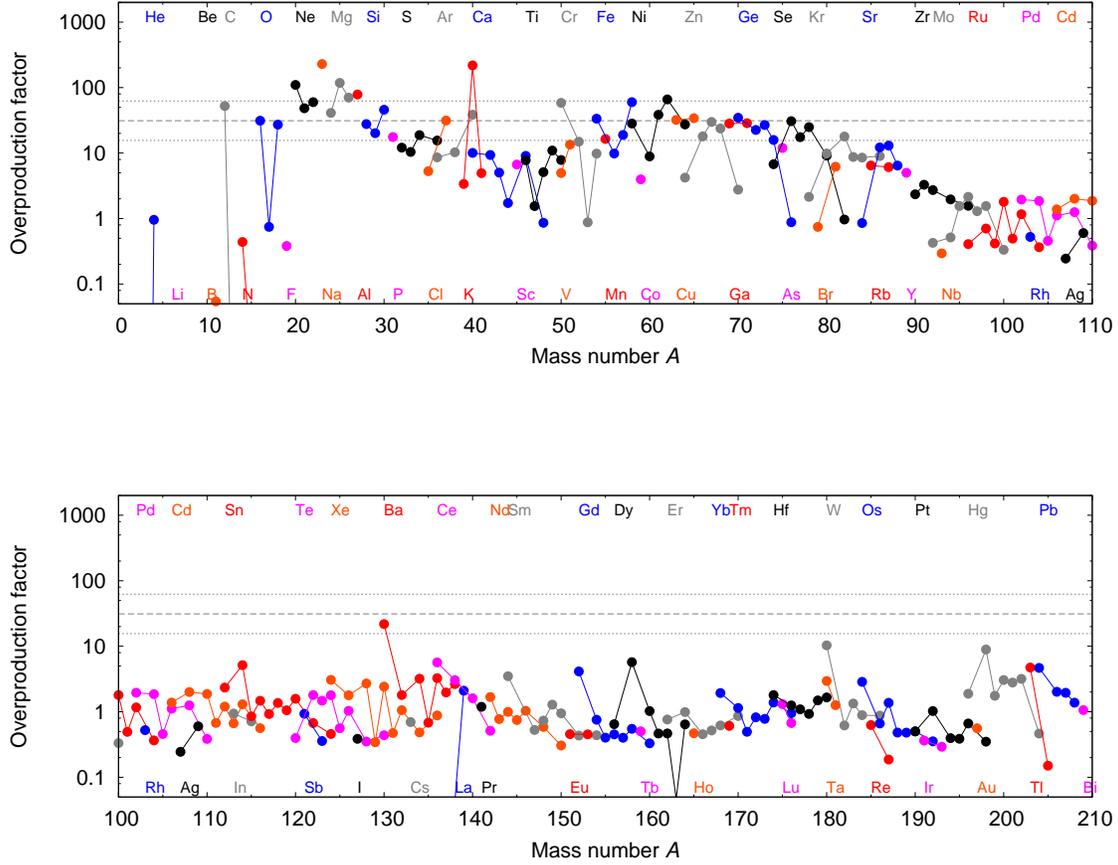}} 
\end{tabular}
\vspace{-1.5cm}
\caption{Same as Fig.~\ref{fig:all_fycf} but for the OKK-CF85 model.}
\label{fig:all_okkcf}
\end{center}
\end{figure}

\begin{figure}
\begin{center}
\begin{tabular}{cc}
\resizebox{15cm}{!}{\includegraphics{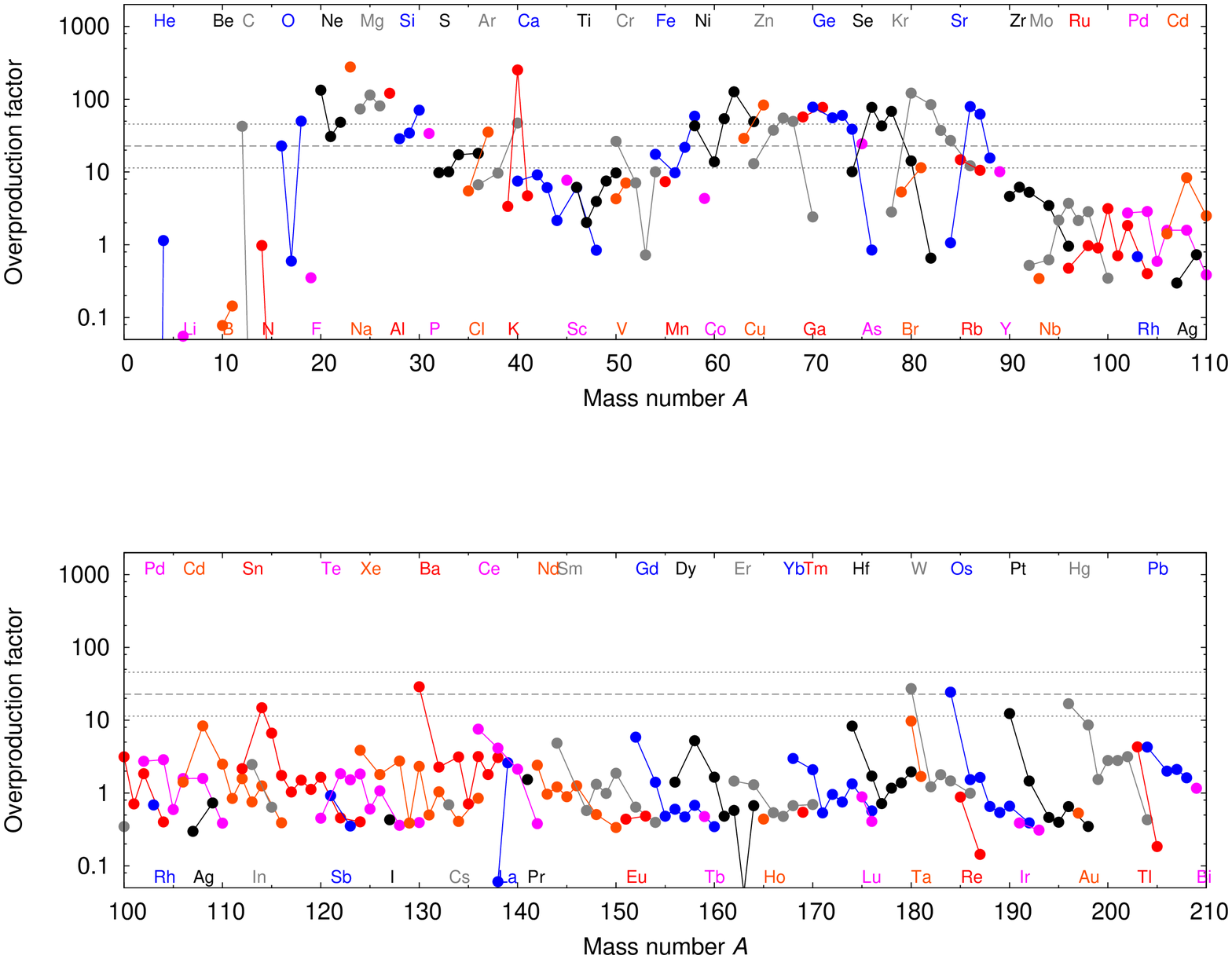}} 
\end{tabular}
\vspace{-1.5cm}
\caption{Same as Fig.~\ref{fig:all_fycf} but for the OKK-Bu96 model.}
\label{fig:all_okkbu}
\end{center}
\end{figure}

To compare the results with observations, 
overproduction factors, $X(i)/X(i)_{\odot}$, are considered. 
We show the results for stable elements lighter than $A$ = 210 in Fig.~\ref{fig:all_fycf} (Fynbo-CF85), Fig.~\ref{fig:all_fybu} (Fynbo-Bu96), Fig.~\ref{fig:all_okkcf} (OKK-CF85) and Fig.~\ref{fig:all_okkbu} (OKK-Bu96).
We classify the produced nuclei into two groups, one is the nuclei produced in mainly massive stars and ejected by the explosions (nuclei of $20 \leq A \leq 32$ and weak $s$-nuclei of $60 < A < 90$) and the other is $p$-nuclei. 
The underproduced nuclei are contributed by other astrophysical sites.
Nuclei of $A < 20$ are also synthesized in low- and intermediate-mass stars. 
The intermediate-mass stars yield main $s$-nuclei ($A = 90-210$) during the AGB star phase \cite{straniero1995,gallino1998,arlandini1999}.
Type Ia supernovae synthesize the nuclei between Cl and Fe group nuclei.  $R$-nuclei could be produced by 
neutron star mergers \cite{wanajo2014,rosswog2014,korobkin2012,goriely2011},  magnetorotationally-driven  supernovae~\cite{nishi06,saru} and/or neutrino driven supernovae \cite{wanajo2013}. 
In the following sections (\S 3.1 and \S 3.2), we discuss the overproduction factors by focusing on different results due to 
four sets of reaction rates.

\subsection{Overproduction factors for $A \leq 110$}
Here, we consider overproduction factors averaged in the ejecta at the time of 10$^{17}$ s after the explosion. 
In Figs.~\ref{fig:all_fycf}-\ref{fig:all_okkbu}, the overproduction factors for Fynbo-CF85, Fynbo-Bu96, OKK-CF85, 
and OKK-Bu96 models are shown. The dotted lines indicate the values whose ratios to the overproduction factor 
of $^{16}$O are two or one-half. If the ratios are found within the values between one-half and two, 
they would be consistent with the solar abundances.
The Fynbo-CF85 model reproduces well the solar system abundances of $20 \leq A < 32$, 
where $^{21}$Ne could be barely in the range. 
For the Fynbo-Bu96 model, isotopes of $20 \leq A < 32$ are well reproduced except for $^{23}$Na and $^{24}$Mg.
For the OKK-CF85 model, $^{20}$Ne, $^{23}$Na, $^{25,26}$Mg, and $^{27}$Al are beyond 
the values whose ratios to that of $^{16}$O are two. 
(these nuclei are synthesized from $^{12}$C+$^{12}$C reactions). 
$^{22}$Ne, which is produced from $^{14}$N, and nuclei of $27 \leq A \leq 32$ are reproduced enough. 
For the OKK-Bu96 model, those of $20 \leq A \leq 27$ are extremely overproduced but $28 \leq A < 32$ well reproduced.

The weak $s$-process produced the elements up to $A = 90$ in all four models. 
Some $s$-nuclei are destroyed by the explosion but almost other $s$-nuclei are survived. 
Therefore, the yields after the explosion are nearly the same as those in presupernova stage.
The overproduction factors for the OKK-CF85 model are the least enhanced among the four models, 
and those of the OKK-Bu96 model are the most enhanced especially around $A = 90$. 
These differences are related to He- and C-burning, which are crucially important for the weak $s$-process~\cite{kiku2012}.
For the OKK-Bu96 model, $^{12}$C is produced appreciably compared to other three models, which leads to the increase in neutron productions during C-burning.

\subsection{Overproduction factors of $p$-nuclei}

In general, $p$-nuclei are produced by way of photodisintegration of seed $s$-nuclei during a supernova explosion.
 It has been found the condition for the adequate $p$-process to occur. The relationship between synthesized $p$-nuclei and the peak temperature 
$T_{\rm p}$ ,maximum temperature at each Lagrange 
mass coordinate during the passage of the supernova shock wave, has been given with the mass number $A$ and neutron number $N$ as follows \cite{phra90,rahpn95}.
\begin{enumerate}
\item $T_{\rm p}=2\times 10^9$ K : Reactions ($\gamma,\,$n)  dominate the $p$-process. small fraction of the heaviest and most fragile seed nuclei $N > 82$ are destroyed and produce heavy $p$-nuclei ($N > 82, A > 140$).
\item $T_{\rm p}=2\times 10^9$ - $2.4 \times 10^9$ K : $(\gamma,\alpha)$ and/or ($\gamma,\,$p) reactions become also active ($N > 82$) and heavy $p$-nuclei are synthesized through the ($\gamma,\,$n) reactions.
\item $T_{\rm p}=2.4$ - $2.6 \times 10^9$ K : $P$-nuclei of $N > 82$ are destroyed by photodisintegration. After the reactions freeze out, the elements become stable nuclei by way of $\beta ^+$ decay. Some intermediate $p$-nuclei are 
produced.
\item $T_{\rm p}= 2.7$ - $3.0 \times 10^9$ K : Seed nuclei of $50 < A < 82$  are photodisintegrated and most intermediate $(50 < N < 82)$ $p$-nuclei are synthesized.
\item $T_{\rm p} > 3\times 10^9$ K : All heavy seed nuclei are destroyed. Light $p$-nuclei ($N < 50$) are produced. 
\end{enumerate}


\begin{figure}
\begin{center}
\begin{tabular}{cc}
\resizebox{14cm}{!}{\includegraphics{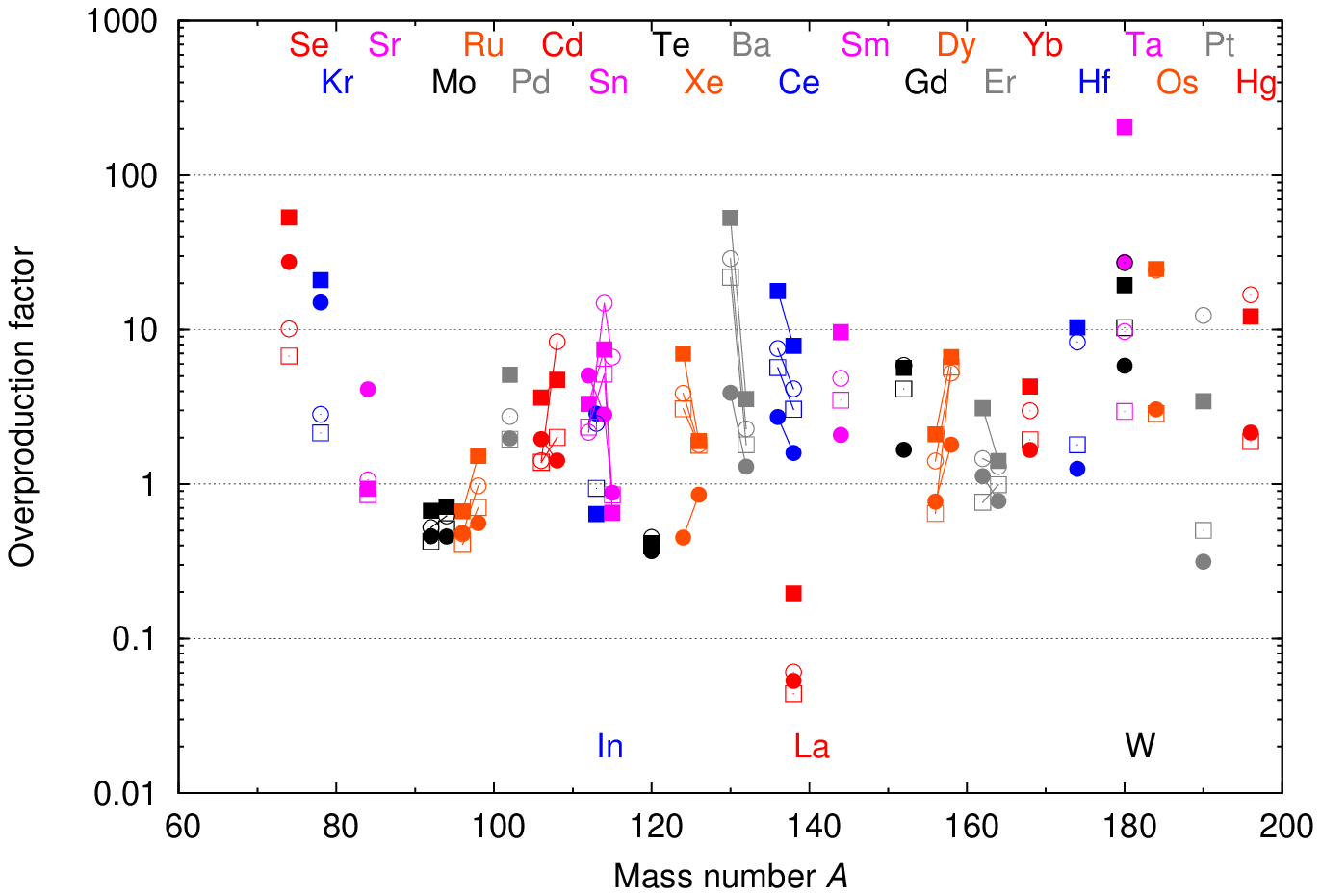}} 
\end{tabular}
\vspace{1.5cm}
\caption{Overproduction factors of all $p$-nuclei against the mass number $A$ at
the time of 10$^{17}$ s after the explosion. All unstable parent nuclei are decayed to stable daughter $p$-nuclei. 
Filled squares are the results of the Fynbo-CF85 model, filled circles are those of Fynbo-Bu96, empty squares are those of OKK-CF85 and empty circles are those of OKK-Bu96.}
\label{fig:p_ex}
\end{center}
\end{figure}
\begin{figure}[b!]
\begin{center}
\begin{tabular}{cc}
\resizebox{12cm}{!}{\includegraphics{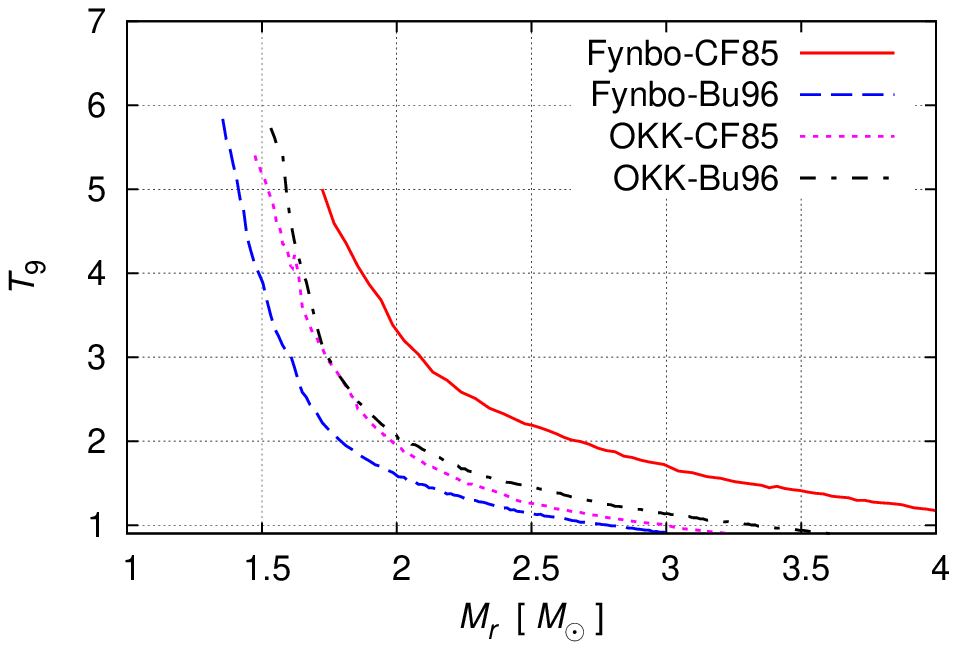}} 
\end{tabular}
\vspace{1.5cm}
\caption{Peak temperatures against mass coordinate $M_r$. $T_9$ denotes $T/(10^9$ K). Solid line is the result of Fynbo-C85 model, dot-dashed line is of OKK-Bu96, dotted line is of OKK-CF85 and deahed line is of OKK-Bu96. 
$P$-process layers (PPLs)~\cite{rahpn95} are defined as the regions with the peak temperature of $(2-3.5) \times 10^9$ K.
}
\label{fig:peak_t}
\end{center}
\end{figure}

In Fig.~\ref{fig:p_ex}, the overproduction factors of all $p$-nuclei after the explosion are plotted.
Most $p$-nuclei are produced in descending order of Fynbo-CF85, OKK-Bu96, OKK-CF85, and Fynbo-Bu96 models.
To consider the reason of the differences among the models seen 
in Fig.~\ref{fig:p_ex}, we take into account the relationship between the $p$-process 
and the peak temperature. 
We define so-called $p$-process layers (hereafter PPLs)~\cite{rahpn95}  as the regions with the peak temperatures of $(2-3.5) \times 10^9$ K. 
Peak temperatures against the Lagrange mass coordinate are shown in Fig.~\ref{fig:peak_t}. 
The size of PPLs for Fynbo-CF85 model is equal to 0.65 $M_\odot$, the amount of which is the largest among the four models. 
Towards the gravitational collapse, the Fynbo-CF85 model forms gradually higher temperature and density regions trough the stellar evolution. 
The sizes of PPLs for other models are 0.27 $M_\odot$ (OKK-Bu96 model), 0.24 $M_\odot$ (OKK-CF85 model) and 0.19 $M_\odot$ (Fynbo-Bu96 model). 
Therefore, the overproductions of $p$-nuclei are in descending order of the sizes of the PPLs. 
Since peak temperatures attained by the shock wave propagation depend on the density distribution 
or the stellar radius at the presupernova stage, 
the amount of $p$-nuclei is affected by the gravitational contraction and/or shell burnings. 
It is noted that  produced  isotopes of $^{92,94}$Mo and $^{96,98}$Ru  are still under-produced; Furthermore, both $^{113}$In and $^{115}$Sn are produced to some extent. These nuclei have been known to be significantly under-produced \cite{rahpn95}. 
Variations of production of $p$-nuclei depend on the survived seed $s$-nuclei.
The difference in overproduction of above $p$-nuclei compared to the previous study \cite{rahpn95} is attributed to the detailed calculations of the nucleosynthesis 
during the stellar evolution. 

\subsection{Summary of nucleosynthesis}

We have shown the supernova nucleosynthesis for the 25$M_{\odot}$ stars 
using the presupernova models which are the results of the stellar evolution calculations with four sets of 3$\alpha$ and 
$^{12}$C($\alpha,\gamma$)$^{16}$O reaction rates and the postprocessing nucleosynthesis with the large nuclear 
reaction netowork. 
We emphasize that final results of supernova nucleosynthesis depend on not only the explosion episode but 
also the history of stellar evolution towards the Fe-core collapse. 
Generally speaking, the models with the OKK rate overproduce the isotopes of Ne, Mg, and Na beyond an acceptable level, 
which are originated from the burning of $^{12}$C and subsequent shell burnings. 
For all models, the amount of $s$-nuclei does not change appreciably compared to that of the presupernova stage. 
As a consequence, He- and C-burnings are significantly important for the weak $s$-process.
On the other hand, the distribution and amount of $p$-nuclei depend on peak temperatures and
 the size of PPLs, which are affected by the stellar evolution path, i.e., 3$\alpha$ and 
$^{12}$C($\alpha,\gamma$)$^{16}$O reaction rates. 
Although each overproduction factors are influenced to some extent,
the heavy element nucleosynthesis is not affected appreciably by
the triple-$\alpha$ and $^{12}$C($\alpha$,$\gamma$)$^{16}$O rates as a whole. 
Therefore, it is difficult to testify the validity of the two reaction rates by the $s$- and $p$-process elements.


\section{Discussions}

We have investigated the effects of 3$\alpha$ and $^{12}\rm C(\alpha,\gamma)^{16}O$ reaction rates on the
production of the supernova yields in a massive star of 25 $M_{\odot}$, where four combinations of the 
representative reaction rates are selected and incorporated in the nuclear reaction network. 
Since the evolutionary code used in the present study is almost the same as Ref.~\cite{rf:nh88,rf:hashi95} 
but for the reaction rates, the differences in the evolutionary path should come from 
the extent of the convective mixing originated from nuclear burnings due to the different reaction rates. 
For example, the stellar evolutionary path of 20 $M_{\odot}$ is affected seriously if we adopt the combination of 
OKK-Bu96 reaction rates, because produced carbon induces the strong carbon-shell burnings. 
Concerning the 25 $M_{\odot}$ star, we can perform the evolutionary calculations till the presupernova stages and 
obtain the Fe-cores just before the collapse for all the combinations of the reaction rates. 
As a consequence, we can recognize significant effects on supernova yields. 1) Distribution of abundance before the core collapse becomes very different for each model. 2)  Supernova explosion results in distinctive yields if we compare them with the solar
system abundances. 
The Fynbo-CF85 model can reproduce the solar values well for $A<40$ 
and it becomes difficult to reproduce the solar ones in ascending order of the 
Fynbo-Bu96, OKK-CF85, and OKK-Bu96 models as seen in Fig. 8--11. 
It should be noted that $^{23}$Na is much overproduced except for 
the Fynbo-CF85 model. Therefore, Fynbo-CF85 
is the most suitable combination of the 3$\alpha$ and 
$^{12}$C($\alpha$,$\gamma$)$^{16}$O reaction rates to be compatible with 
the solar system abundances. It is noted that the CF85 rate for 
$^{12}$C($\alpha$,$\gamma$)$^{16}$O reaction is considered to be the upper limit 
within the experimental uncertainties. 

As for the heavy nuclei beyond iron group elements, it is unclear how to judge the compatibility with the observations. 
It remains the problem of underproduction of $p$-nuclei compared to the solar values~\cite{rayet1990}.
There exists crucial problems concerning the stellar models of which we do not have a satisfactory theory of convective mixing.
Since we have adopted the Schwartzschild criterion for convection, the convection tends to occur rather easily 
compared to the Leudux criterion. Furthermore, the extent of convective mixing is not known well, 
where convection itself is related to nuclear burnings closely~\cite{smith+arnett14}. 
We should note that these problems arise from the assumption that the stars are spherically symmetric and at present any satisfactory calculation of non-spherical stellar evolution does not exist~\cite{kippen,maeder}. 
Although the helium core is assumed to be 8 $M_{\odot}$, the actual star
begins the evolution from the main-sequence stage with the hydrogen rich envelope. If we consider the hydrogen-rich envelope, we
will worry about the mass loss rate which brings out uncertain parameters \cite{smith2014}.
As far as the approach of the helium star is concerned, our results would be legitimate because, after the end of hydrogen burning, a star forms the helium core with clear boundary between the core and envelope, which is equivalent to the helium star~\cite{rf:sn80}.
Observationally, our approach has been supported from the explanation of light curves~\cite{snhs87,snh88} and 
from the supernova nucleosynthesis~\cite{ha1989,rf:hashi95} as far as SN 1987A is concerned.
Therefore, at present our conclusion could be accepted even if the unsatisfactory theory of convection lies under the calculation
of stellar evolution.

\section*{Acknowledgment}
We thank to Dr. Masahiro Machida, Takashi Teranishi  and Kenzo Arai for helpful discussions.
This work has been supported in part by a Grant-in-Aid for Scientific Research (24540278) of the Ministry of Education, Culture, Sports, Science and Technology of Japan.



\begin{thebibliography}{99}
\bibitem{rf:sn80}D. Sugimoto and K. Nomoto, \JL{Space Sci. Rev.,25,155,1980}.
\bibitem{rf:nh88}K. Nomoto and M. Hashimoto, \JL{Phys. Rep.,163,13,1988}.
\bibitem{rf:hashi95}M. Hashimoto, \JL{Prog. Theor. Phys.,94,663,1995}.
\bibitem{rf:nom82}K. Nomoto, \JL{Astrophys. J.,253,798,1982}.
\bibitem{rf:nom82b}K. Nomoto, \JL{Astrophys. J.,257,780,1982}.
\bibitem{rf:ntm85}K.~Nomoto, F.-K.~Thielemann, and S.~Miyaji, \JL{Astron. Astrophys.,149,239,1985}.
\bibitem{rf:okk}K. Ogata, M. Kan, and M. Kamimura, \JL{Prog. Theor. Phys.,122,1005,2009}.
\bibitem{rf:angulo}C. Angulo et al. \JL{Nucl. Phys. A,656,3,1999}.
\bibitem{rf:fynbo}H. O. U. Fynbo et al., \JL{Nature,433,136,2005}.
\bibitem{Fowler1975} W. A. Fowler, G. R. Caughlan, and B. A. Zimmerman, \JL{Annual Review of Astron. Astrophys.,13,69,1975}.
\bibitem{cf88}G. R. Caughlan and W. A. Fowler, \JL{Atomic Data and Nuclear Data Tables,40,283,1988}.
\bibitem{cfhz85}G. R. Caughlan, W. A. Fowler, M. J. Harris, and B. A. Zimmerman, \JL{Atomic Data and Nuclear Data Tables,32,197,1985}.
\bibitem{bu96}L. Buchmann, R.E. Azuma, C.A. Barnes, J. Humblet, and K. Langanke, \JL{Phys. Rep. C,54,354,1996}.
\bibitem{tur2007} C. Tur, A. Heger, and S. M. Austin, \JL{Astrophys. J.,671,821,2007} 
\bibitem{tur2009}C. Tur, A. Heger, and S. M. Austin, \JL{Astrophys. J.,702,1068,2009}.
\bibitem{austin2014} S. M. Austin, C. West, and A. Heger, \JL{Phys. Rev. Lett.,112,111101,2014}. 
\bibitem{parikh2013} A. Parikh, J. Jose, I. R. Seitenzahl, and F. K. Ropke, \JL{Astron. Astrophys.,557,A3,2013}. 
\bibitem{rf:dotter}A. Dotter and B. Paxton, \JL{Astron. Astrophys.,507,1617,2009}.
\bibitem{rf:morel}P. Morel, J. Provost, B. Pichon, Y. Lebreton, and F. Th${\rm {\acute e}}$venin, \JL{Astron. Astrophys.,520,A41,2010}.
\bibitem{suda}T. Suda, R. Hirschi, and M. Y. Fujimoto1, \JL{Astrophys. J.,741,61,2011}.
\bibitem{mm97}G. Meynet and A. Maeder, \JL{Astron. Astrophys.,321,465,1997}.
\bibitem{maeder97}A. Maeder, \JL{Astron. Astrophys.,321,134,1997}.
\bibitem{meakin+arnett07}C. A. Meakin and D. Arnett, \JL{Astrophys. J.,667,448,2007}.
\bibitem{smith+arnett14}N. Smith and W. D. Arnett, \JL{Astrophys. J.,785,82,2014}.
\bibitem{rf:piot07}G. Piotto et al., \JL{Astrophys. J.,661,L53,2007}.
\bibitem{saru2010}M. Saruwatari and M. Hashimoto, \JL{Prog. Theor. Phys.,124,925,2010}.
\bibitem{matsuo2011}Y. Matsuo et al., \JL{Prog. Theor. Phys.,126,1177,2011}.
\bibitem{rayet00} M. Rayet and M. Hashimoto, \JL{Astron. Astrophys.,354,740,2000}.
\bibitem{phn90} N. Prantzos, M. Hashimoto, and K. Nomoto, \JL{Astron. Astrophy., 234,211,1990}.
\bibitem{ono2012} M. Ono, M. Hashimoto, S. Fujimoto, K. Kotake, and S. Yamada, \JL{Prog. Theor. Phys.,128,4,2012}.
\bibitem{kiku2012} Y. Kikuchi, M. Ono, Y. Matsuo, M. Hshimoto, and S. Fujimoto, \JL{Prog. Theor. Phys.,127,171,2012}.
\bibitem{nishi06} S. Nishimura, et al., \JL{Astrophys. J.,642,410,2006}.
\bibitem{cyburt} R. H. Cyburt et al., \JL{Astrophys. J.,189,240,2010}.
\bibitem{bao00} Z. Y. Bao et al., \JL{Atomic Data and Nucl. Data Tables,76,70,2000}.
\bibitem{ty87} K. Takahashi and K. Yokoi, \JL{Atomic Data and Nucl. Data Tables,36,375,1987}.
\bibitem{un2002}H. Umeda and K. Nomoto, \JL{Astrophys. J.,565,385,2002}.
\bibitem{nh86}K. Nomoto and M. Hahimoto, \JL{Prog. Part. Nucl. Phys.,17,267,1986}.
\bibitem{latter91}W. B. Latter, \JL{Astrophys. J.,377,187,1991}.
\bibitem{gh97}J. S. Greaves and W. S. Holland, \JL{Astron. Astrophys.,327,342,1997}.
\bibitem{ha1989}M. Hashimoto, K. Nomoto, and T. Shigeyama, \JL{Astron. Astrophys.,210,L5,1989}.
\bibitem{yahili} M. D. Johnston and A. Yahil \JL{Astrophys. J.,285,587,1984}.
\bibitem{snh88} T. Shigeyama, K. Nomoto, and M. Hashimoto, \JL{Astron. Astrophy.,196,141,1988}.
\bibitem{straniero1995}  O. Straniero, R. Gallino, M. Busso, A. Chieffl, C. M. Raiteri, M. Limongi, and M. Salaris, \JL{Astrophys. J.,440,L85,1995}. 
\bibitem{gallino1998} R. Gallino, C. Arlandini, M. Busso, M. Lugaro, C. Travaglio, O. Straniero, A. Chieffl, and M. Limongi, \JL{Astrophys. J.,497,388,1998}. 
\bibitem{arlandini1999} C. Arlandini, F. Kappeler, K. Wisshak, R. Gallino, M. Lugaro, M. Busso, and O. Straniero, \JL{Astrophys. J.,525,886,1999}. 
\bibitem{wanajo2014} S. Wanajo, Y. Sekiguchi, N. Nishimura, K. Kiuchi, K. Kyutoku, and M. Shibata, \JL{Astrophys. J.,789,L39,2014}. 
\bibitem{rosswog2014} S. Rosswog, O. Korobkin, A. Arcones, F.-K. Thielemann, and T. Piran, \JL{Mon. Not. Roy. Astron. Soc.,439,744,2014}. 
\bibitem{korobkin2012} O. Korobkin, S. Rosswog, A. Arcones, and C. Winteler, \JL{Mon. Not. Roy. Astron. Soc.,426,1940,2012}. 
\bibitem{goriely2011} S. Goriely, A. Bauswein, and H.-T. Janka, \JL{Astrophys. J.,738, L32,2011}. 
\bibitem{saru} M. Saruwatari, M. Hashimoto, R. Fukuda, and S. Fujimoto, \JL{Journal of Astrophys.,2013,506146,2013}.
\bibitem{wanajo2013} S. Wanajo, \JL{Astrophys. J.,770, L22,2013}. 
\bibitem{rahpn95} M. Rayet, M. Arnould, M. Hashimoto, N. Prantzos, and K. Nomoto, \JL{Astron. Astrophys.,298,517,1995}.
\bibitem{phra90} N. Prantzos, M. Hashimoto, M. Rayet, and M. Arnould, \JL{Astron. Astrophy.,238, 455,1990}.
\bibitem{rayet1990} M. Rayet, N. Prantzos, and M. Arnould, \JL{Astron. Astrophys.,227,271,1990}. 
\bibitem{koike1999} O. Koike, M. Hashimoto, K. Arai, and S. Wanajo, \JL{Astron. Astrophys.,342,464,1999}.
\bibitem{kippen} R. Kippenhahn and A. Weigert, Stellar Structure and Evolution (Springer-Verlag, 199).
\bibitem{maeder} A. Maeder, Physics, Formation and Evolution of Rotating Stars (Springer, 2009).
\bibitem{smith2014} N. Smith, \JL{Annual Review of Astronomy and Astrophysics,52,487,2014}. 
\bibitem{snhs87} T. Shigeyama, K. Nomoto, M. Hashimoto, and D. Sugimoto, \JL{Nature,328,320,1987}.



       








\end{thebibliography}
%

\vfill\pagebreak

\end{document}